\documentclass[a4paper]{PoS}
\usepackage{graphicx}
\usepackage{amsmath}
\usepackage{amssymb}
\usepackage{color}
\usepackage{cite}
\usepackage{verbatim}
\usepackage{setspace}
\usepackage{xspace}
\usepackage{slashed}
\usepackage{fontenc}
\usepackage{times}
\usepackage{mathptmx}
\usepackage[small]{subfigure}
\usepackage{bm}
\newcommand{\nn}{\nonumber}

\newcommand{\taubar}{\bar\tau}
\newcommand{\eps}{\epsilon}
\newcommand{\bfT}{{\mathbf T}}
\def\rd{\mathrm{d}}

\title{Automated Calculation of ${\pmb N}$-jet Soft Functions}

\ShortTitle{Automated Calculation of $N$-jet Soft Functions}

\author{Guido Bell\\
        Theoretische Physik 1, Naturwissenschaftlich-Technische Fakult\"at, Universit\"at Siegen,\\
        Walter-Flex-Strasse 3, 57068 Siegen, Germany\\
        E-mail: \email{bell@physik.uni-siegen.de}}

\author{\speaker{Bahman Dehnadi}\\
     Theoretische Physik 1, Naturwissenschaftlich-Technische Fakult\"at, Universit\"at Siegen,\\
     Walter-Flex-Strasse 3, 57068 Siegen, Germany\\
     E-mail: \email{dehnadi@physik.uni-siegen.de}}
        
\author{Tobias Mohrmann\\
       Theoretische Physik 1, Naturwissenschaftlich-Technische Fakult\"at, Universit\"at Siegen,\\
       Walter-Flex-Strasse 3, 57068 Siegen, Germany\\
        E-mail: \email{mohrmann@physik.uni-siegen.de}} 

\author{Rudi Rahn\\
        Albert Einstein Center for Fundamental Physics, Institut f\"ur Theoretische Physik,\\ 
        Universit\"at Bern, Sidlerstrasse 5, 3012 Bern, Switzerland\\
        E-mail: \email{rahn@itp.unibe.ch}} 
        
\abstract{ 
We present a systematic framework for the calculation of soft functions that are defined in 
terms of $N\geq2$ light-like Wilson lines. The formalism represents an extension of a method 
that we developed earlier for the calculation of dijet soft functions to the general $N$-jet 
case. We discuss the technical aspects of this generalisation, focussing on SCET-1 soft 
functions that obey the non-Abelian exponentiation theorem in this contribution. As a first 
application of our method, we consider the $N$-jettiness observable and present numerical 
results for the $1$-jettiness and $2$-jettiness hadron-collider soft functions to 
next-to-next-to-leading order in the perturbative expansion.}

\FullConference{Loops and Legs in Quantum Field Theory (LL2018)\\
		29 April 2018 - 04 May 2018\\
		St. Goar, Germany}

\begin{document}
%%%%%%%%%%%%%%%%%%%%%%%%%%%%%%%%%%%%%%%%%%%%%%%%%%%%%%%%%%%%%%%%%%%%%%

\section{Introduction}	

Soft functions are essential ingredients of factorisation theorems that arise whenever the QCD 
radiation to a hard-scattering process is restricted to the soft and collinear regions. 
Factorisation is the basis for an all-order resummation of logarithmic corrections, which can 
be achieved e.g.~by solving renormalisation group (RG) equations in Soft-Collinear Effective 
Theory (SCET)~\cite{Bauer:2000yr,Bauer:2001yt,Beneke:2002ph}. In the past years, the 
factorised cross sections have also been used as a subtraction (or slicing) technique 
for fixed-order QCD calculations, with the $q_T$ subtraction~\cite{Catani:2007vq} and the 
$N$-jettiness subtraction~\cite{Boughezal:2015dva,Gaunt:2015pea} being the most prominent 
examples.

In this contribution we present a generalisation of a method that we developed previously 
for the calculation of soft functions that are defined in terms of two back-to-back light-like 
Wilson lines~\cite{Bell:2015lsf,Bell:2018jvf,Bell:2018vaa}. As a first step to the general 
$N$-jet case, we focus in this work on SCET-1 soft functions that obey the non-Abelian 
exponentiation (NAE) theorem~\cite{Gatheral:1983cz,Frenkel:1984pz}. In the following, 
we discuss the technical aspects of our approach, and present numerical results for the 
$N$-jettiness soft function to next-to-next-to-leading order (NNLO) in the perturbative 
expansion.

\section{$\pmb N$-jet soft functions}
\label{sec:N-jet-soft-function}

We consider soft functions of the generic form
\begin{equation}
\label{eq:N-jet-soft-func}
S(\tau, \mu) = \sum_X \; 
\mathcal{M}(\tau;\lbrace k_{i} \rbrace)\;
\langle 0 | \big(S_{n_1} S_{n_2}\cdots S_{n_N}\big)^\dagger| X \rangle
\langle X | S_{n_1} S_{n_2}\cdots S_{n_N} | 0 \rangle \,,
\end{equation}
where $S_{n_i}$ are soft Wilson lines extending along $N\geq2$ light-like directions 
$n_i^\mu$. The Wilson lines depend on the colour representation of the associated hard partons, 
and the soft function itself is a matrix in colour space. The path of the Wilson lines 
-- and hence the $i\epsilon$ prescription of the eikonal propagators -- furthermore differs for 
incoming and outgoing partons. The term $\mathcal{M}(\tau;\lbrace k_{i} \rbrace)$ 
represents a generic measurement function that provides a constraint on the soft radiation with 
momenta $\lbrace k_{i} \rbrace$. The specific form we assume for the measurement function will 
be given in the following sections, but we generally assume that it is defined in Laplace space 
with $\tau$ denoting the corresponding Laplace variable.

Assuming that the tree-level (no emission) measurement function is normalised to one, the 
perturbative expansion of the $N$-jet soft function can be written in the form
\begin{equation}
S(\tau, \mu) = 1 + \left(\frac{Z_\alpha\alpha_s}{4\pi}\right) (\mu^2 \taubar^2)^\eps \; 
S^{(1)}(\eps) + \left(\frac{Z_\alpha\alpha_s}{4\pi}\right)^2 (\mu^2 \taubar^2)^{2\eps} \; 
S^{(2)}(\eps) + \mathcal{O}(\alpha_s^3)\,,
\end{equation}
where $\eps=(4-d)/2$ is the dimensional regulator, $\taubar = \tau e^{\gamma_E}$ and 
$\alpha_s$ is the $\overline{\textrm{MS}}$-renormalised coupling constant, which is related 
to the bare coupling constant $\alpha_s^0$ via 
$Z_\alpha \alpha_s\,\mu^{2\eps}=e^{-\eps\gamma_E}(4\pi)^\eps \alpha_s^0$ with 
$Z_\alpha = 1-\beta_0\alpha_s/(4\pi\eps)$ and $\beta_0 = 11/3\, C_A - 4/3\,T_F n_f$. 
In this notation, the first term represents the unit operator in colour space, whereas the 
NLO and NNLO coefficients $S^{(1,2)}(\eps)$ have a non-trivial colour structure, and they 
implicitly depend on the kinematic factors $ n_{ij}\equiv n_i\cdot n_j$. Our strategy for 
computing these coefficients closely follows the one we adopted for the calculation of dijet 
soft functions in~\cite{Bell:2015lsf}. In particular, we assume in this work that the 
soft functions are not sensitive to rapidity divergences (and are therefore of SCET-1 type) 
and that they are consistent with NAE.

\section{NLO calculation}
\label{sec:N-jettiness-NLO}

At NLO the virtual corrections are scaleless and vanish in dimensional regularisation. The 
real-emission diagrams that connect the same Wilson line also vanish, since the reference 
vectors $n_i^\mu$ are all light-like. We are thus left at this order with real-emission 
diagrams that connect a pair of different Wilson lines, and the soft function is thus given 
by a sum over dipole contributions,
\begin{equation}
S^{(1)}(\eps) = \sum_{i\neq j} \, \bfT_{i} \cdot \bfT_{j} \,
\Big( \frac{n_{ij}}{2} \Big)^\eps \, S_{ij}^{(1)}(\eps)\,,
\end{equation}
where $\bfT_i$ and $\bfT_j$ are the colour generators of the $i$'th and $j$'th parton in the
colour-space notation of~\cite{Catani:1996vz}. 
The dipole contributions can be calculated in close analogy to the method we developed for 
dijet soft functions in~\cite{Bell:2015lsf}. We thus start from
\begin{equation}
S_{ij}^{(1)}(\eps)  = -\frac{(2\pi\, n_{ij}\,e^{\gamma_E} \tau^2)^{-\eps}}{(2\pi)^{d-1}} \,
\int d^{d}k \;\, \delta(k^{2}) \,\theta(k^{0}) 
\,\mathcal{M}_1(\tau; k) \, |\mathcal{A}_{ij}(k)|^{2} \,,
\end{equation}
where $|\mathcal{A}_{ij}(k)|^{2}=16 \pi^2 n_{ij}/k_ik_j$ is the square of the dipole matrix 
element with $k_i = k\cdot n_i$ and $k_j = k\cdot n_j$. We furthermore decompose the gluon 
momentum according to\\[-0.0mm]
\begin{equation}
k^\mu =  k_j \,\frac{n_i^\mu}{n_{ij}} +\, k_i \,\frac{n_j^\mu}{n_{ij}} + k_{\perp}^{\mu}\,,
\end{equation}
where $k_{\perp}^{\mu}$ is a vector that is transverse to the jet directions $n_i^\mu$ and 
$n_j^\mu$. In $d$ dimensions the transverse space can be parametrised by $(d-2)$ components, 
and we find it convenient to work in a frame in which the dipole jets are back-to-back, such 
that the transverse space is purely space-like and can be expressed using ordinary spherical 
coordinates. We further introduce the boost-invariant variables 
$ k_{T}= \sqrt{2 k_i k_j/n_{ij}}$ and $y=k_i/k_j$, and write the one-emission measurement 
function as
\begin{equation}
\label{eq:measure:NLO}
\mathcal{M}_1(\tau; k) = \exp\bigg\{-\tau\, \sqrt{\frac{n_{ij}}{2}} \,k_{T}\bigg( 
y^{n/2}\, f_A(y,\theta_1,\theta_2) \,\theta(1-y) + 
y^{-n/2}\, f_B(1/y,\theta_1,\theta_2) \,\theta(y-1)
\bigg)\bigg\}\,,
\end{equation}
which consists of two hemisphere contributions for $y\leq1$ and $y\geq1$. In the first term,
we factor out an appropriate power of the rapidity variable $y$ to make sure that the function 
$f_A(y,\theta_1,\theta_2)$ is finite and non-zero in the limit $y\to0$, and we do so similarly 
for the second term for $y\to\infty$~\cite{Bell:2015lsf}. The exponential typically arises 
from a Laplace transformation, and we assume that the Laplace variable $\tau$ has dimension 
$1/$mass, which fixes the linear dependence on $k_T$ on dimensional grounds. We further account 
for two angular dependences, since the projection of the gluon momentum onto the jet 
directions $n_k^\mu$ can in general be expressed in terms of two angles in the transverse 
plane~\cite{BDMR}.

After performing the observable-independent integrations and mapping $y\to 1/y$ in the second 
hemisphere contribution, we obtain the following representation of the NLO dipole contribution,
\begin{align}
\label{eq:NLO:master}
S_{ij}^{(1)}(\eps)  &= \frac{2\,e^{-\gamma_E\eps}}{\pi}\,
\frac{\Gamma(-2\eps)}{\Gamma(-\eps)}\;\int_0^1 dy \;\, y^{-1+n\eps}\;
\int_{-1}^1 d\cos\theta_1 \;\,\sin^{-1-2\eps}\theta_1  \nn\\[0.2em]
&\qquad\quad
\int_{-1}^1 d\cos\theta_2 \;\,\sin^{-2-2\eps}\theta_2\;\;
\Big\{f_A(y,\theta_1,\theta_2)^{2\eps}
+ f_B(y,\theta_1,\theta_2)^{2\eps}\Big\}\,.
\end{align}
Notice that for SCET-1 soft functions with $n\neq0$, the rapidity integral produces a 
pole in $\eps$ in the collinear limit $y\to0$. The factor $\Gamma(-2\epsilon)$ furthermore
captures a soft singularity that arises in the limit $k_T\to 0$. We also remark that the 
integration over the angle $\theta_2$ produces an unphysical (spurious) divergence, which 
is compensated by the prefactor $1/\Gamma(-\epsilon)\sim \mathcal{O}(\epsilon)$. Our result
in \eqref{eq:NLO:master} generalises the corresponding expression for dijet soft functions
in~\cite{Bell:2015lsf}, where we assumed that the two jets are back-to-back and that the 
observable is symmetric under $n\leftrightarrow\bar n$ exchange.

\section{NNLO calculation}
\label{sec:N-jettiness-NNLO}

At NNLO the two-loop virtual corrections are again scaleless and vanish. The remaining 
contributions are the mixed real-virtual (RV) and double real-emission corrections. The 
latter consists of two contributions: the emission of a soft quark-antiquark 
pair ($q\bar q$) and the emission of two soft gluons ($gg$). Hence the total NNLO correction 
to the soft function can be written in the form
\begin{equation}
S^{(2)}(\eps)= S^{(2,{\rm RV})}(\eps) +{S}^{(2,q\bar{q})}(\eps) +{S}^{(2,gg)}(\eps)\,.
\end{equation}
The real-virtual correction consists of two-particle and three-particle 
correlations~\cite{Catani:2000pi},
\begin{equation}
\label{eq:NNLO:RV}
S^{(2,{\rm RV})}(\eps) =  C_A \,\sum_{i\neq j} \, \bfT_{i} \cdot \bfT_{j} \,
\Big( \frac{n_{ij}}{2} \Big)^{2\eps} \, S_{ij}^{(2,{\rm Re})}(\eps)
+ \sum_{i\neq j\neq k} (\lambda_{ij} - \lambda_{ip}-\lambda_{jp})\,
 f_{ABC}\; \bfT_{i}^A \;\bfT_{j}^B \;\bfT_{k}^C\;
 S_{ijk}^{(2,{\rm Im})}(\eps)\,,
\end{equation}
where $\lambda_{AB}=1$ if partons $A$ and $B$ are both incoming or outgoing, and 
$\lambda_{AB}=0$ otherwise. Here the indices $i,j,k$ refer to the hard partons associated
with the respective Wilson lines and $p$ represents the emitted soft gluon. The first term 
again has a dipole structure and the corresponding matrix element is proportional to
$ (n_{ij}/2k_{i}k_{j})^{1+\epsilon}$. The second term reveals a three-parton colour 
correlation, which we will refer to as a tripole contribution. It arises from the imaginary 
part of the loop integral, and it is thus process dependent. In processes with three 
hard partons, one can show that the tripole contribution vanishes because of colour 
conservation~\cite{Catani:2000pi}, but for general processes with four or more partons 
it leads to a non-trivial correction. The corresponding matrix element is of the form
$ (n_{ij}/2k_{i}k_{j})^{\epsilon} (n_{ik}/2k_{i}k_{k})$. As both matrix elements
resemble the one of the NLO calculation, we apply a similar phase-space 
parametrisation to factorise the divergences.

For the double real-emission contribution, we obtain~\cite{Catani:1999ss}
\begin{align}
{S}^{(2,q\bar{q})}(\eps) &= 
 T_F \,n_f \,\sum_{i\neq j} \, \bfT_{i} \cdot \bfT_{j} \,
\Big( \frac{n_{ij}}{2} \Big)^{2\eps}  S_{ij}^{(2,q\bar{q})}(\eps)\,,
 \\ 
{S}^{(2,gg)}(\eps) &= 
 C_A \,\sum_{i\neq j} \, \bfT_{i} \cdot \bfT_{j} \,
\Big( \frac{n_{ij}}{2} \Big)^{2\eps} S_{ij}^{(2,gg)}(\eps)
+ \frac{1}{4} \, \sum_{i \neq j} \sum_{k \neq l}
\big\{\bfT_{i} \cdot \bfT_{j}\,,\,\bfT_{k} \cdot \bfT_{l}\big\} 
\Big( \frac{n_{ij}\,n_{kl}}{4} \Big)^{\eps}
%\Big( \frac{n_{ij}}{2} \Big)^{\eps}\Big( \frac{n_{kl}}{2} \Big)^{\eps}
S_{ij}^{(1)}(\eps)  \,S_{kl}^{(1)}(\eps),\nn
\end{align}
where in the last term we asumed that the observable is consistent with NAE, and
\begin{equation}
S_{ij}^{(2,X)}(\eps) = -
\frac{(2\pi\, n_{ij}\,e^{\gamma_E} \tau^2)^{-2\eps}}{(2\pi)^{2d-2}} \,
\int d^{d}k \;\, \delta(k^{2}) \,\theta(k^{0}) \,
\int d^{d}l \;\, \delta(l^{2}) \,\theta(l^{0}) 
\;\mathcal{M}_2(\tau; k,l) \, |\mathcal{A}_{ij}^X(k,l)|^{2} 
\end{equation}
for $X\in\{q\bar{q},gg\}$. The corresponding matrix elements for both colour structures 
$|\mathcal{A}_{ij}^X(k,l)|^{2}$ can be found in~\cite{Catani:1999ss}. Unlike the 
single-emission case, the double real-emission correction exhibits overlapping 
divergences, which we disentangle with a suitable phase-space reparametrisation~\cite{BRT}.

For the double real-emission contribution, we again follow~\cite{Bell:2015lsf} and 
introduce the variables
\begin{equation}
p_T = \sqrt{\frac{2(k_i+l_i)(k_j+l_j)}{n_{ij}}} \,, \qquad
y = \frac{k_i+l_i}{k_j+l_j} \,, \qquad
a = \sqrt{\frac{k_{j}\,l_{i}}{k_{i}\,l_{j}}} \,,\qquad
b =\sqrt{\frac{k_{i}\,k_{j}}{l_{i}\,l_{j}}} \,.
\end{equation}
In terms of these variables, our ansatz for the two-emission measurement function becomes
\begin{align}
\label{eq:measure:NNLO}
\mathcal{M}_2(\tau; k,l) &= \exp\bigg\{-\tau\, \sqrt{\frac{n_{ij}}{2}} \,p_{T}\,
\theta(1-a)\\
&\quad\times
\bigg( 
y^{n/2}\Big[ F_A\big(a,b,y,\{\theta_i\}\big) \,\theta(1-b)
+ F_B\big(a,1/b,y,\{\theta_i\}\big) \,\theta(b-1)\Big]\theta(1-y)\nn\\
&\qquad
+ 
y^{-n/2}\Big[ F_C\big(a,b,1/y,\{\theta_i\}\big) \theta(1-b)
+ F_D\big(a,1/b,1/y,\{\theta_i\}\big) \theta(b-1)\Big]\theta(y-1)
\bigg)\bigg\},\nn
\end{align}
where  $\{\theta_i\} \equiv \{\theta_{k_1},\theta_{k_2},\theta_{k_3},\theta_{l_1},\theta_{l_2}\}$
represents a set of angles that we use to express the momenta of the two emitted partons in the 
transverse space~\cite{BDMR}. Furthermore, similar to the single-emission case in
\eqref{eq:measure:NLO}, we have factorised the linear dependence on $p_T$ on dimensional 
grounds, as well as the asymptotic behaviour of the measurement function in the limits $y\to0$ 
and $y\to\infty$. We further exploit the symmetry under $k\leftrightarrow l$ exchange, to map
the contribution from $a\geq1$ onto $a\leq1$. The functions $F_i$ satisfy certain constraints
from infrared and collinear safety, which we discussed in detail for the dijet case
in~\cite{Bell:2018vaa}. With the phase-space parametrisation and the measurement function 
at hand, one can then derive a similar master formula for the double real-emission contribution 
as in the NLO case, in which all singularities are factorised. Further details will 
be given in~\cite{BDMR}.

\section{Renormalisation}
\label{sec:RGE}

In Laplace space the relation between the bare and the renormalised soft function reads
\begin{equation}
\label{eq:Sbare}
 S_0(\tau) =  Z_S(\tau, \mu) \; S(\tau, \mu)\; Z_S^{\dagger}(\tau, \mu)\,,
\end{equation}
where the counterterm $Z_S(\tau, \mu)$ is a matrix in colour space. The $\overline{\text{MS}}$ 
renormalised soft function satisfies the RG equation
\begin{equation}
\frac{\rd}{\rd \ln\mu}  \; S(\tau,\mu)
= \frac{1}{2} \,\Gamma_S(\tau,\mu) \, S(\tau,\mu) 
+\frac{1}{2} \,S(\tau,\mu)\,  \Gamma_S(\tau,\mu)^{\dagger}
\end{equation}
with anomalous dimension
\begin{equation}
 \Gamma_S(\tau,\mu)= \frac{1}{n}\, \bigg\{
\sum_{i\neq j} \, \bfT_{i} \cdot \bfT_{j} \; \Gamma_{\rm cusp}(\alpha_s)
\bigg[2 \,\ln \bigg(\sqrt{\frac{n_{ij}}{2}}\, \mu \taubar\bigg) 
- i \pi\, \lambda_{ij}\bigg] + 2 \gamma^S(\alpha_s)\bigg\}\,,
\end{equation}
where $\lambda_{ij}$ has been defined after \eqref{eq:NNLO:RV}. The imaginary part is 
related to the anomalous dimension of the associated hard function in the factorisation 
theorem, whose general structure was discussed in~\cite{Becher:2009qa}. The cusp anomalous 
dimension has a perturbative expansion 
$\Gamma_{\rm cusp} (\alpha_s)= \sum_{n=0}^{\infty} \Gamma_n ( \frac{\alpha_s}{4 \pi})^{n+1}$,
with leading coefficients $ \Gamma_0 =4$ and 
$\Gamma_1 = ( 268/9 - 4\pi^2/3) C_A - 80/9\, T_F n_f$.
Up to the considered two-loop order, the soft anomalous dimension also has a dipole 
structure and can be written in the form
$\gamma^{S}(\alpha_s) = \sum_{i\neq j} \, \bfT_{i} \cdot \bfT_{j} \,
\big\{\gamma^{(0)}_{ij} \,(\frac{\alpha_s}{4\pi})
+ \gamma^{(1)}_{ij} \,(\frac{\alpha_s}{4\pi})^2\big\}$.
Following~\cite{Bell:2015lsf,Bell:2018vaa}, we find it furthermore convenient to define 
the anomalous dimensions with a common prefactor $1/n$, where $n$ is related to the 
asymptotic behaviour of the observable in the collinear limit, see \eqref{eq:measure:NLO} and 
\eqref{eq:measure:NNLO}.

The two-loop solution of the RG equation takes the form
\begin{align}
\label{eq:RGE:softsolution}
S(\tau,\mu) &= 
1 + \left( \frac{\alpha_s}{4 \pi} \right) 
\sum_{i\neq j} \, \bfT_{i} \cdot \bfT_{j} \,
\bigg( \frac{\Gamma_0}{n} \,L_{ij}^2 
+ \frac{2\gamma^{(0)}_{ij}}{n} \,L_{ij} 
+ c_{ij}^{(1)} \bigg) 
+\left( \frac{\alpha_s}{4 \pi} \right)^2
\\
&
\times \bigg\{
\sum_{i\neq j} \, \bfT_{i} \cdot \bfT_{j} \,
\bigg(\frac{2\beta_0\Gamma_0}{3n} \,L_{ij}^3 
 + \bigg( \frac{\Gamma_1}{n} + \frac{2\beta_0 \gamma^{(0)}_{ij}}{n}  \bigg) L_{ij}^2 
+ 2 \bigg( \frac{\gamma^{(1)}_{ij}}{n} +\beta_0 c_{ij}^{(1)} \bigg) L_{ij} 
+ c_{ij}^{(2)} \bigg)
\nn\\
&\hspace{7mm}
-2\pi \sum_{i\neq j\neq k} f_{ABC}\; \bfT_{i}^A \;\bfT_{j}^B \;\bfT_{k}^C\;
\bigg( \frac{\lambda_{ij}\Gamma_0}{n^2} \bigg(
\frac{\Gamma_0}{3} L_{jk}^3 +  \gamma^{(0)}_{jk} L_{jk}^2 + n \,c_{jk}^{(1)} L_{jk} \bigg)
+ c_{ijk}^{(2)} \bigg)
\nn\\
&\hspace{7mm}
+ \frac{1}{4} \, \sum_{i \neq j} \sum_{k \neq l}
\big\{\bfT_{i} \cdot \bfT_{j}\,,\,\bfT_{k} \cdot \bfT_{l}\big\} 
\bigg( \frac{\Gamma_0}{n} \,L_{ij}^2 + \frac{2\gamma^{(0)}_{ij}}{n} \,L_{ij} + c_{ij}^{(1)} \bigg) 
\bigg( \frac{\Gamma_0}{n} \,L_{kl}^2 + \frac{2\gamma^{(0)}_{kl}}{n} \,L_{kl} + c_{kl}^{(1)} \bigg)
\bigg\}, \nn
\end{align}
where $L_{ij}= \ln \big(\sqrt{n_{ij}/2}\, \mu \taubar\big)$ etc. Here the 
tripole structure arises from commutators of colour generators, 
e.g.~$\big[ \bfT_{i} \cdot \bfT_{j} \,,\, \bfT_{j} \cdot \bfT_{k} \big] 
= - i f_{ABC} \; \bfT_{i}^A \;\bfT_{j}^B \;\bfT_{k}^C$.

The counterterm, on the other hand, fulfills the RG equation
\begin{equation}
\frac{\rd}{\rd \ln\mu}  \; Z_S(\tau,\mu)
= - \frac{1}{2}  \, Z_S(\tau,\mu) \,\Gamma_S(\tau,\mu)\,,
\end{equation}
and its explicit solution to two-loop order is given by
\begin{align} 
Z_{S}(\tau,\mu) &= 
1 + \left( \frac{\alpha_s}{4 \pi} \right) 
\sum_{i\neq j} \, \bfT_{i} \cdot \bfT_{j} \,
\bigg( \frac{\Gamma_0}{4n} \,\frac{1}{\eps^2}
+ \frac{G^{(0)}_{ij}}{4n} \,\frac{1}{\eps} \bigg) 
+\left( \frac{\alpha_s}{4 \pi} \right)^2 
\\
&\quad
\times \bigg\{
\sum_{i\neq j} \, \bfT_{i} \cdot \bfT_{j} \,
\bigg( -\frac{3\beta_0\Gamma_0}{16n} \,\frac{1}{\eps^3}
+\frac{\Gamma_1-2\beta_0 G^{(0)}_{ij}}{16n} \,\frac{1}{\eps^2}
+ \frac{G^{(1)}_{ij}}{8n} \,\frac{1}{\eps} \bigg) 
\nn\\
&\hspace{11mm}
+ \frac{1}{4} \, \sum_{i \neq j} \sum_{k \neq l}
\big\{\bfT_{i} \cdot \bfT_{j}\,,\,\bfT_{k} \cdot \bfT_{l}\big\} 
\bigg( \frac{\Gamma_0}{4n} \,\frac{1}{\eps^2}
+ \frac{G^{(0)}_{ij}}{4n} \,\frac{1}{\eps} \bigg) 
\bigg( \frac{\Gamma_0}{4n} \,\frac{1}{\eps^2}
+ \frac{G^{(0)}_{kl}}{4n} \,\frac{1}{\eps} \bigg) 
\bigg\}, \nn
\end{align}
where $G^{(k)}_{ij} = (2  L_{ij}  - i \pi \lambda_{ij})\Gamma_{k} +2\,\gamma^{(k)}_{ij}$\,.
Using these expressions, one can reconstruct all divergences of the bare soft function through 
NNLO via \eqref{eq:Sbare}.

\section{$\pmb N$-jettiness soft function}
\label{sec:num-result}

\begin{figure}[t!]
\centering{
 \includegraphics[width=0.32\textwidth]{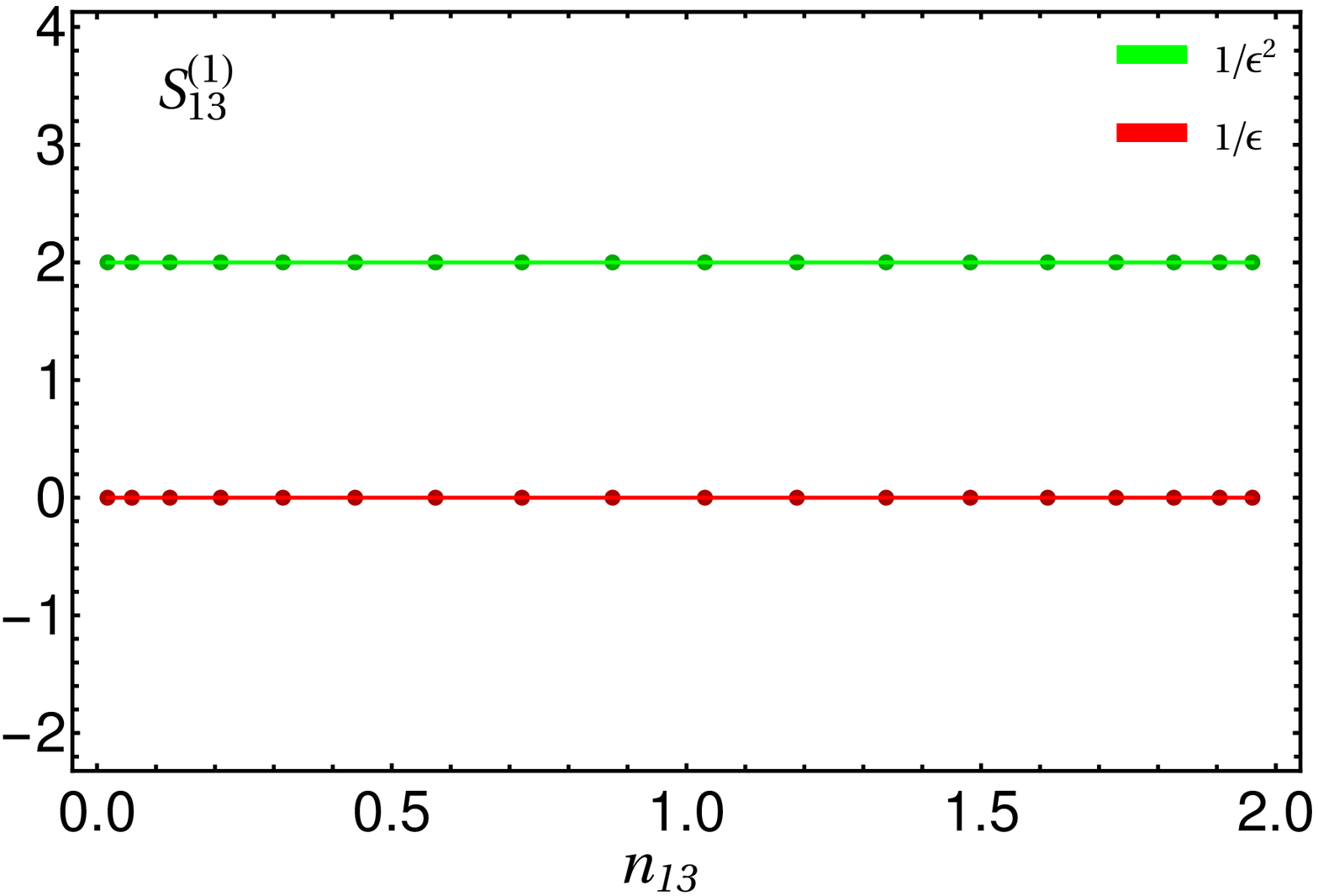}\hspace{2mm}
 \includegraphics[width=0.32\textwidth]{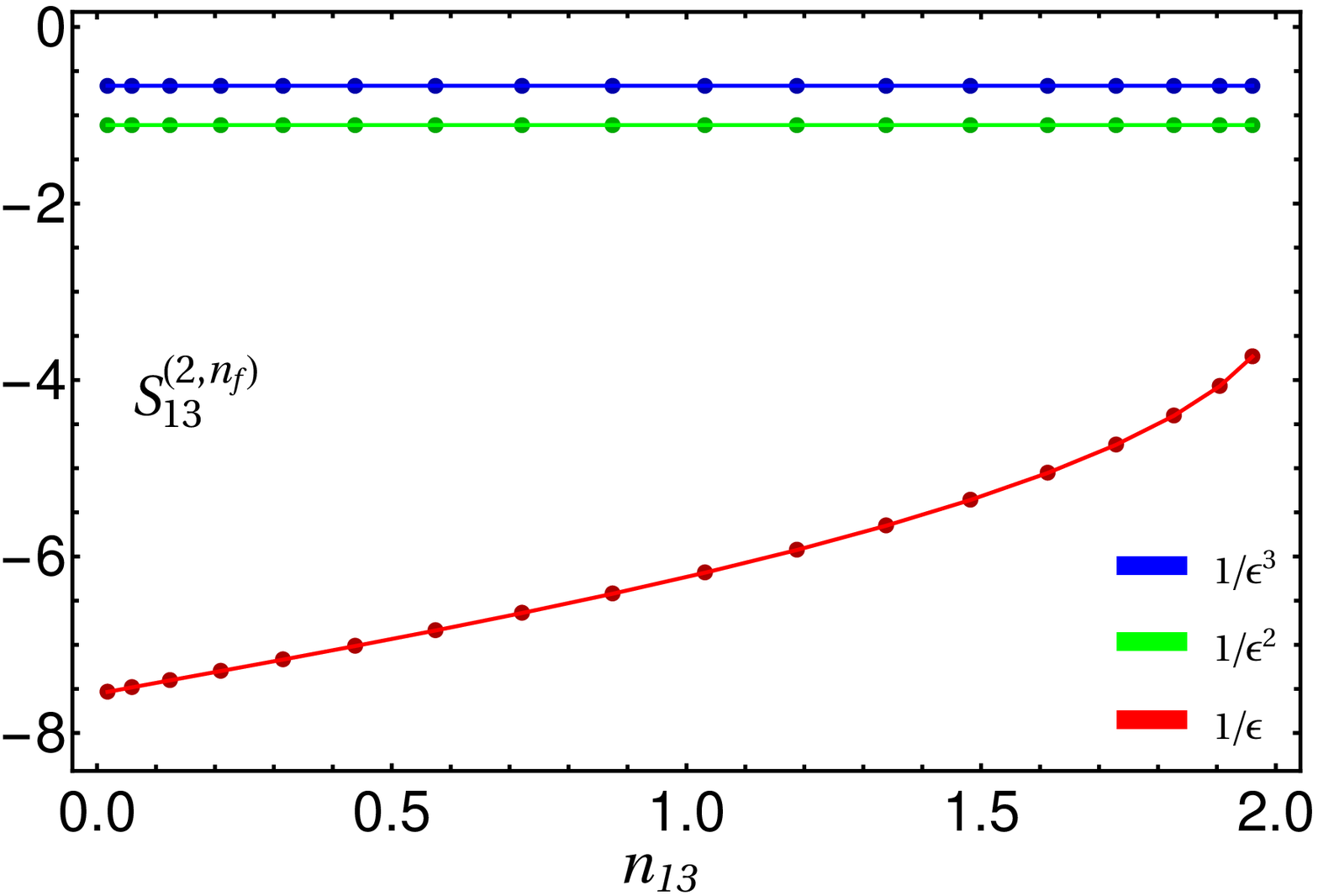}\hspace{2mm}
 \includegraphics[width=0.32\textwidth]{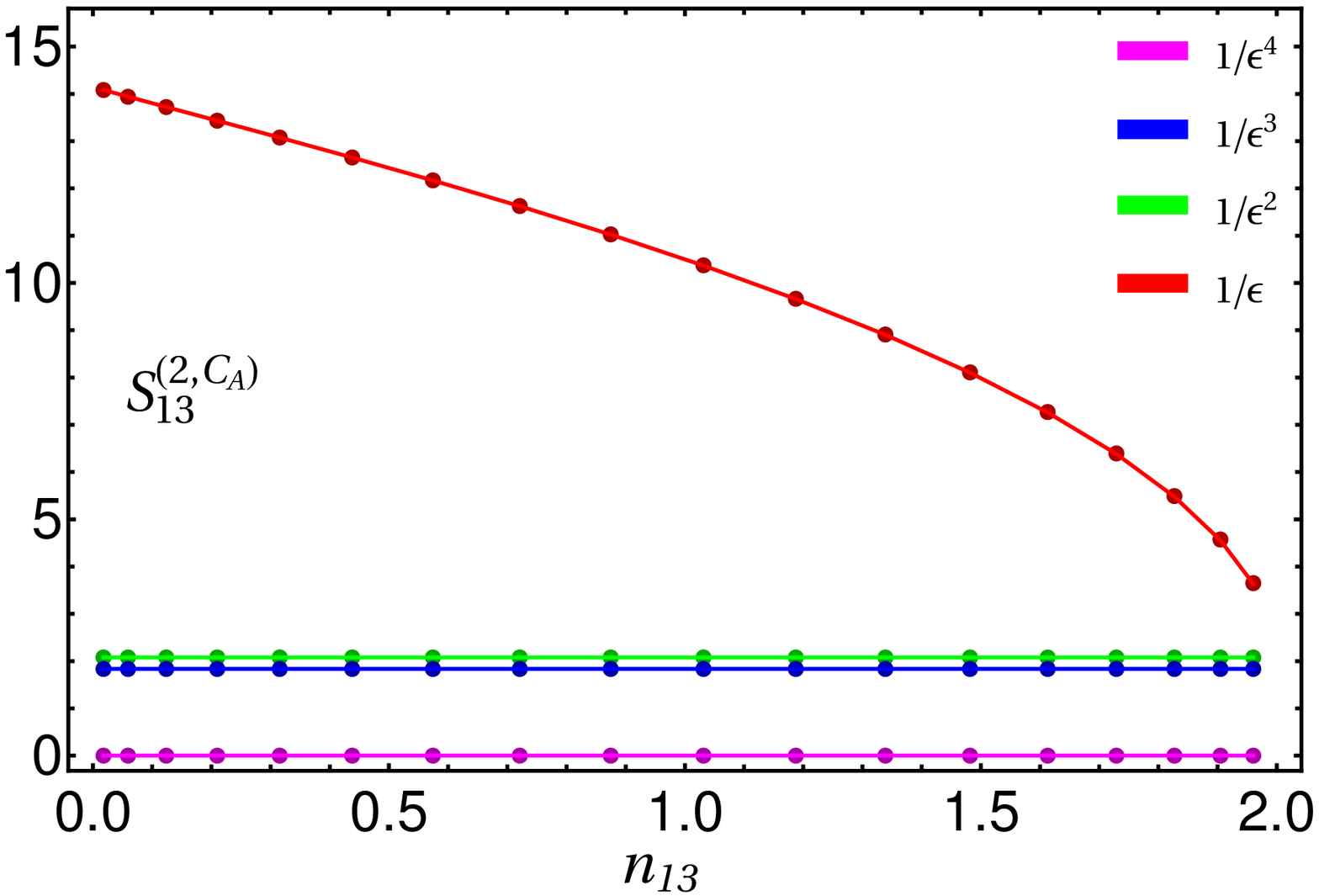}
 \\[0.5em]
 \includegraphics[width=0.32\textwidth]{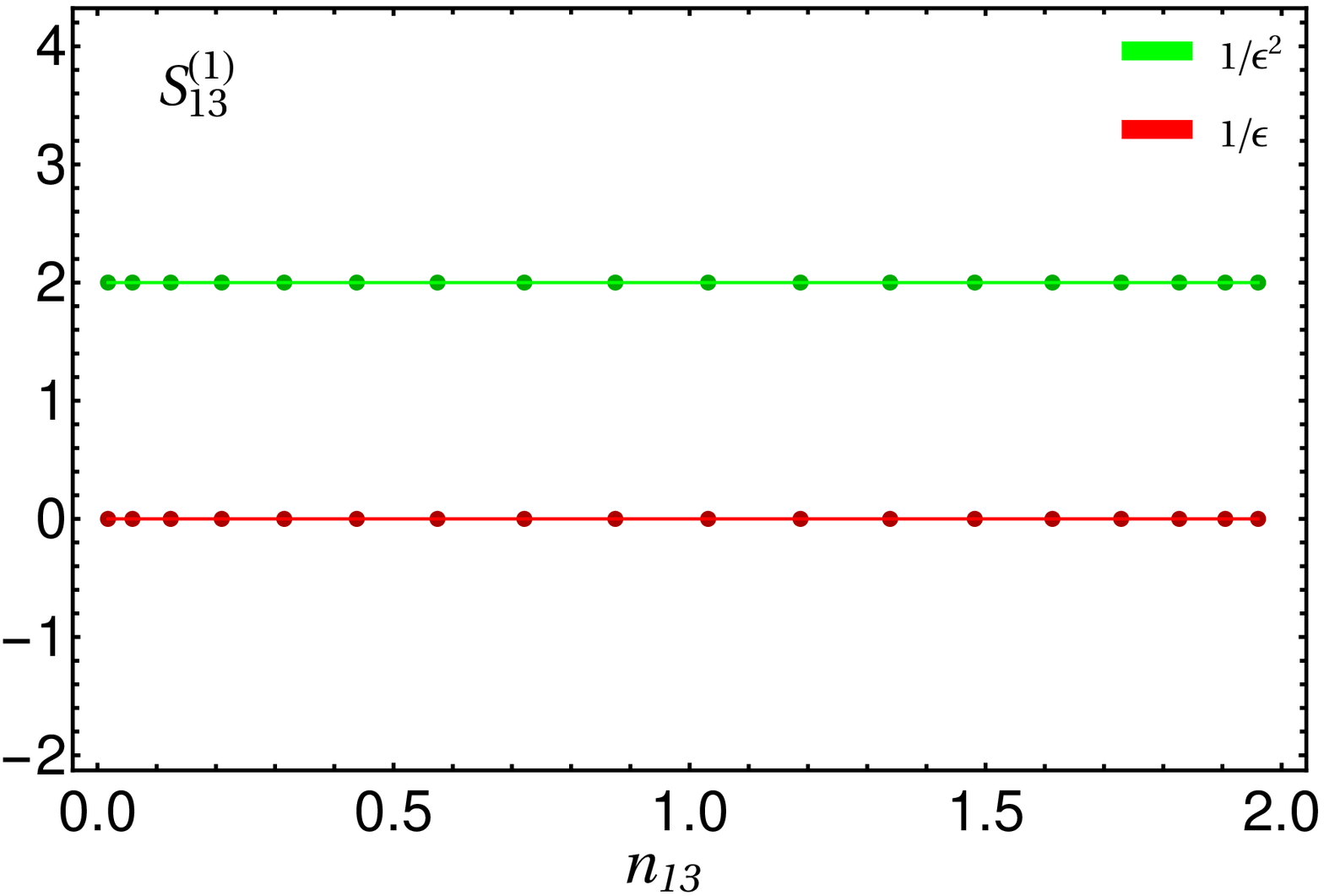}\hspace{2mm}
 \includegraphics[width=0.32\textwidth]{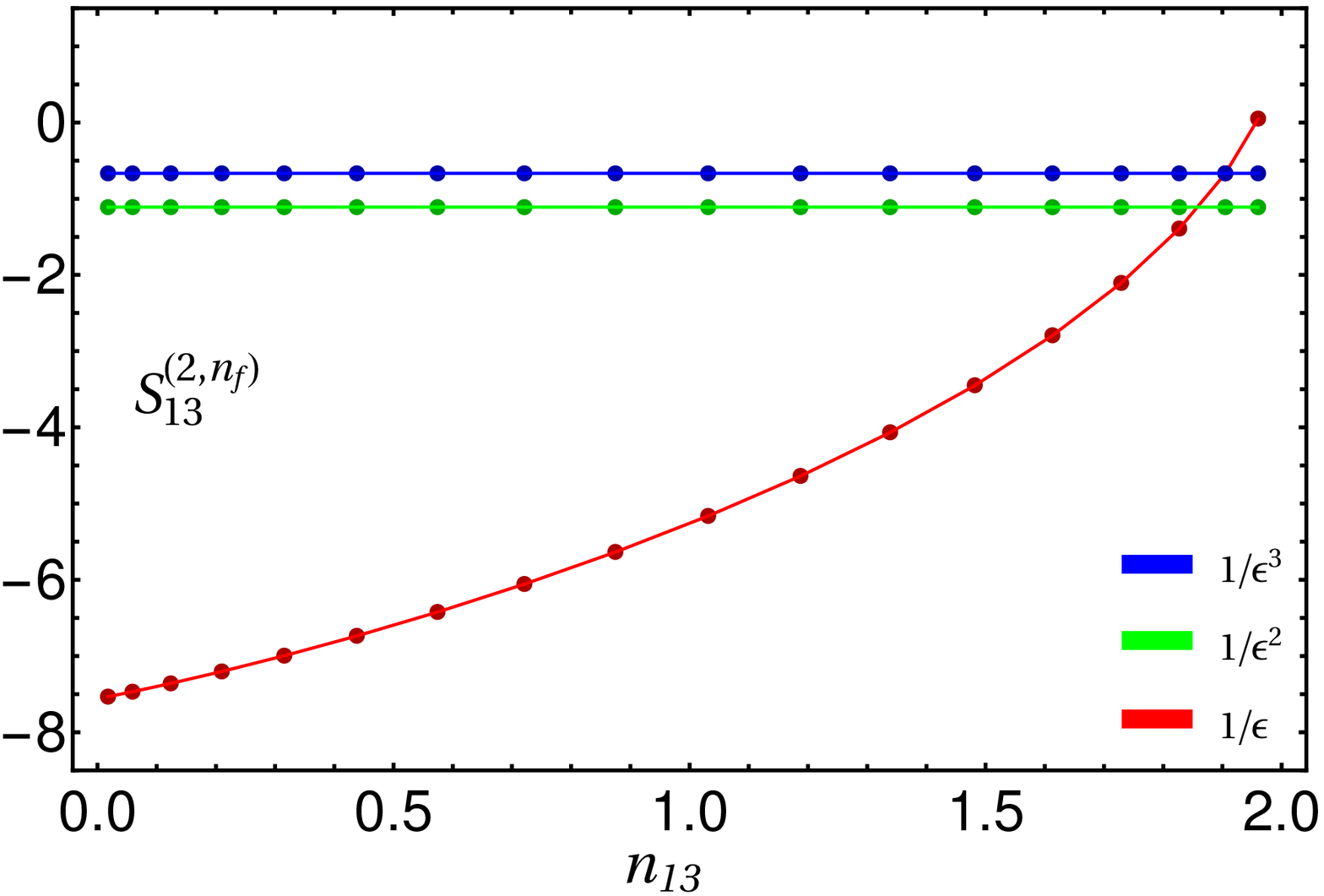}\hspace{2mm}
 \includegraphics[width=0.32\textwidth]{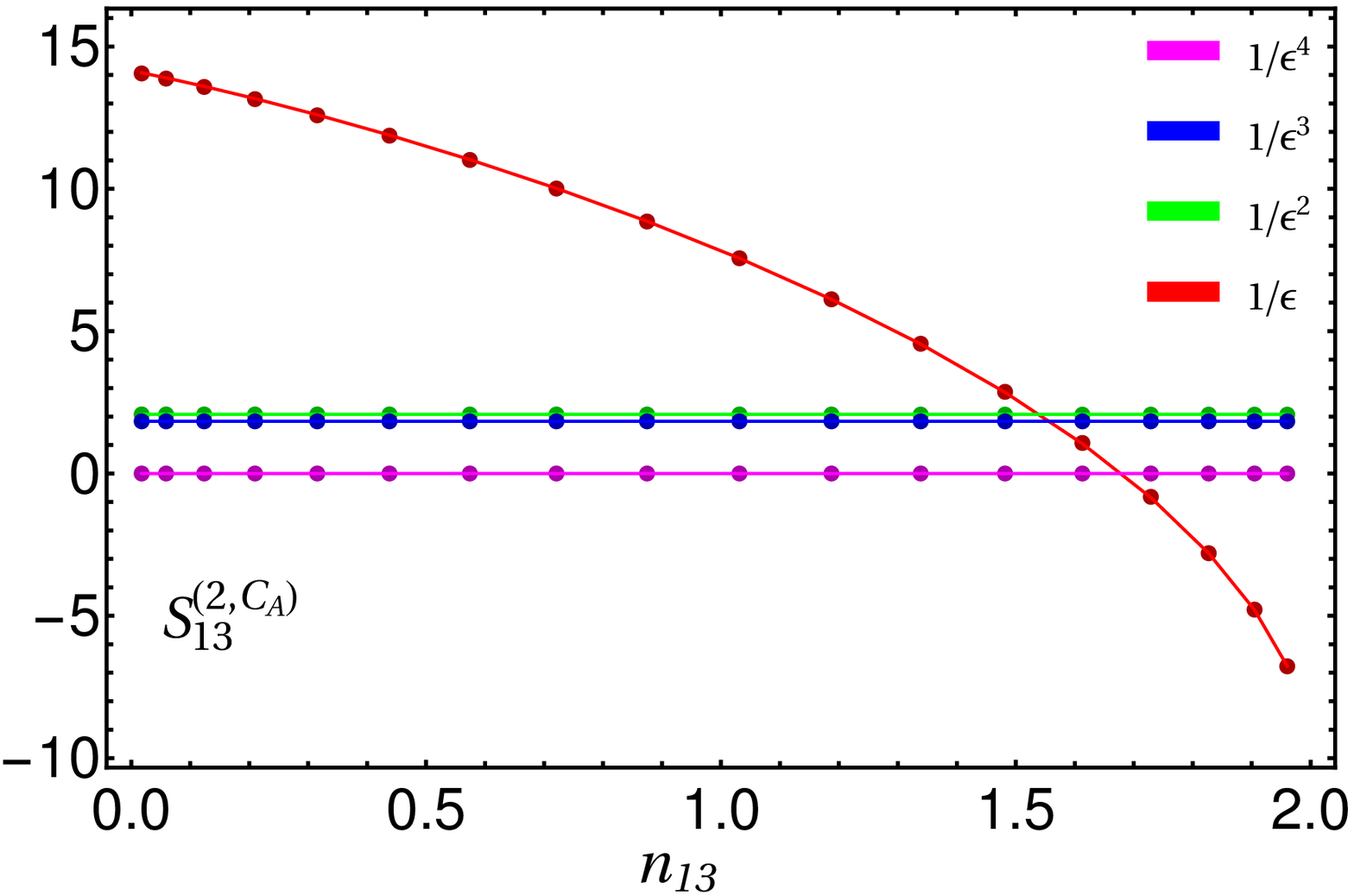}
 } 
 \vspace{-5mm}
\caption{Dipole contributions to the 1-jettiness (upper row) and 2-jettiness (lower row) soft functions. 
The first, second and third column show the $\mathcal{O}(\alpha_s)$, $\mathcal{O}(T_f\,n_f\,\alpha_s^2)$ and 
$\mathcal{O}(C_A\,\alpha_s^2)$ contributions as defined in the text. The dots show our numerical results
for the poles of the bare soft function, and the solid lines are the predictions from the RG equation.}
\label{fig:dipole:poles}
\end{figure}

As an application of our framework, we consider the $N$-jettiness observable at hadron colliders. 
$N$-jettiness is defined as~\cite{Stewart:2010tn}
\begin{equation}
\mathcal{T}_N = \sum_k \; \min_i\big\{n_i \cdot p_k\big\}\,,
\end{equation}
where for simplicity we adopt a specific normalisation that was also used 
in~\cite{Boughezal:2015eha,Campbell:2017hsw}. Here $p_k^\mu$ are the momenta of the final-state radiation, 
and $n_i^\mu$ are a set of light-like reference vectors, with $i=1,2$ denoting the beam directions and 
$i=3,\ldots,(N+2)$ refering to the directions of the final-state jets. The $N$-jettiness hadron collider 
soft function thus corresponds to a $(N+2)$-jet soft function in the terminology of our paper.

The Laplace-space $N$-jettiness soft function falls into the class \eqref{eq:N-jet-soft-func}, and it is 
defined in \mbox{SCET-1} (with parameter $n=1$) and obeys NAE. We can thus apply the developed formalism to 
compute the $N$-jettiness soft function for -- in principle -- arbitrary values of $N$. In the following, 
we show explicit results for the 1-jettiness and 2-jettiness soft functions. As the 1-jettiness soft 
function is already known to NNLO~\cite{Boughezal:2015eha,Campbell:2017hsw}, we can test our method by 
comparing our results to these calculations. The 2-jettiness soft function, on the other hand, is currently 
only known to NLO~\cite{Jouttenus:2011wh}, and our results represent the first NNLO calculation for this 
observable.

We first illustrate that the divergences of the bare soft function calculation agree with the predictions 
from the RG equation that we discussed in Section~\ref{sec:RGE}. The relevant anomalous dimensions 
are all known to the considered order, and can be found in~\cite{Jouttenus:2011wh}. In 
Figure~\ref{fig:dipole:poles} we show the $(13)$-dipole contribution to the 1-jettiness (upper row) and 
2-jettiness (lower row) soft functions for illustration. The plots show the NLO dipole $S_{13}^{(1)}$ 
(left) and the NNLO dipole contributions $S_{13}^{(2,n_f)}=S_{13}^{(2,q\bar q)}$ (middle) and 
$S_{13}^{(2,C_A)}=S_{13}^{(2,{\rm Re})}+S_{13}^{(2,gg)}$ (right) as a function of 
$n_{13}$~\footnote{The 1-jettiness depends on one scattering angle, which can be expressed in terms of 
$n_{13}$. For the 2-jettiness we obtain results for arbitrary kinematics, but for the purpose of 
illustration we assume that the two jets are back-to-back.}. The dots 
represent the numbers that we obtain by processing the formulae from Sections~\ref{sec:N-jettiness-NLO} 
and~\ref{sec:N-jettiness-NNLO} through {\tt pySecDec}~\cite{Borowka:2017idc}\footnote{As the divergences are
completely factorised in our setup, we do not need to perform any sector decomposition step, but 
we rather use {\tt pySecDec} as an interface to the Cuba library.}. 
Our predictions use the Vegas integrator provided by the Cuba library~\cite{Hahn:2004fe}, and they
typically have numerical uncertainties at the subpercent level which are not visible on the scale of the
plots. The numbers of the bare soft function calculation are then compared to the predictions of the 
RG equation, which are illustrated by the solid lines. As can be seen from the plots, the agreement
is very satisfactory for all pole coefficients, which similarly holds for the other dipole contributions.

\begin{figure}[t!]
\centering{
 \includegraphics[width=0.31\textwidth]{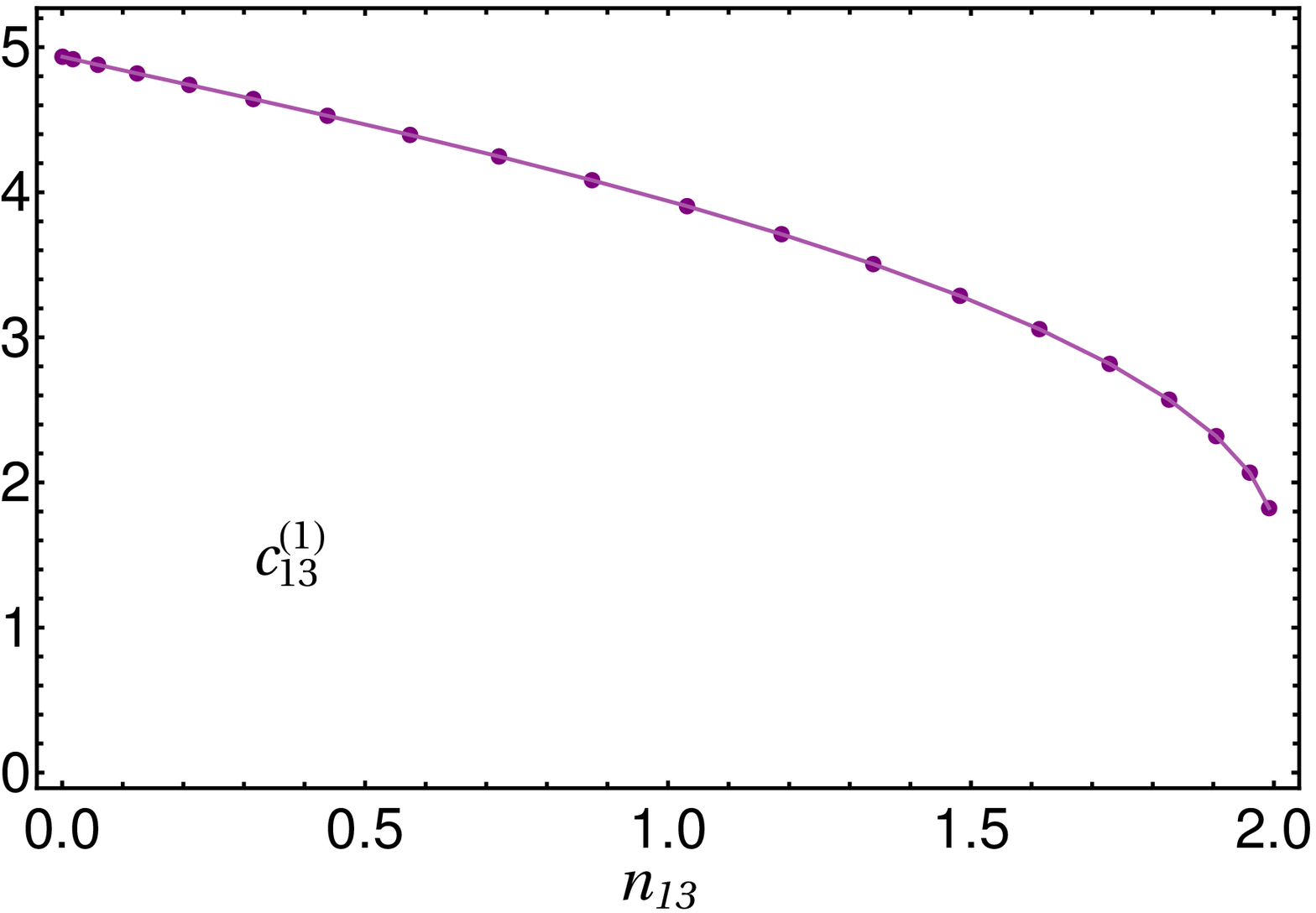}\hspace{2mm}
 \includegraphics[width=0.32\textwidth]{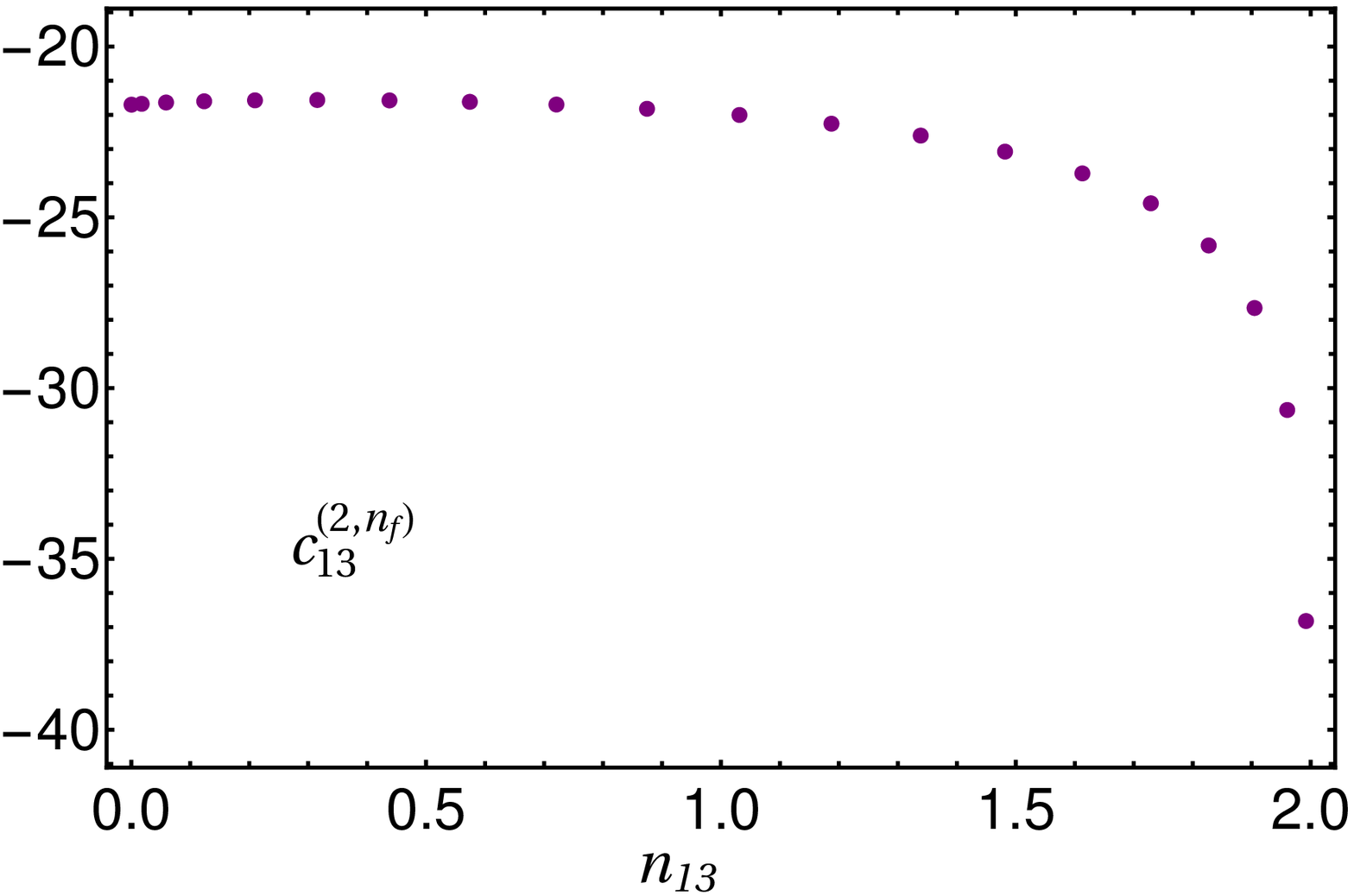}\hspace{2mm}
 \includegraphics[width=0.32\textwidth]{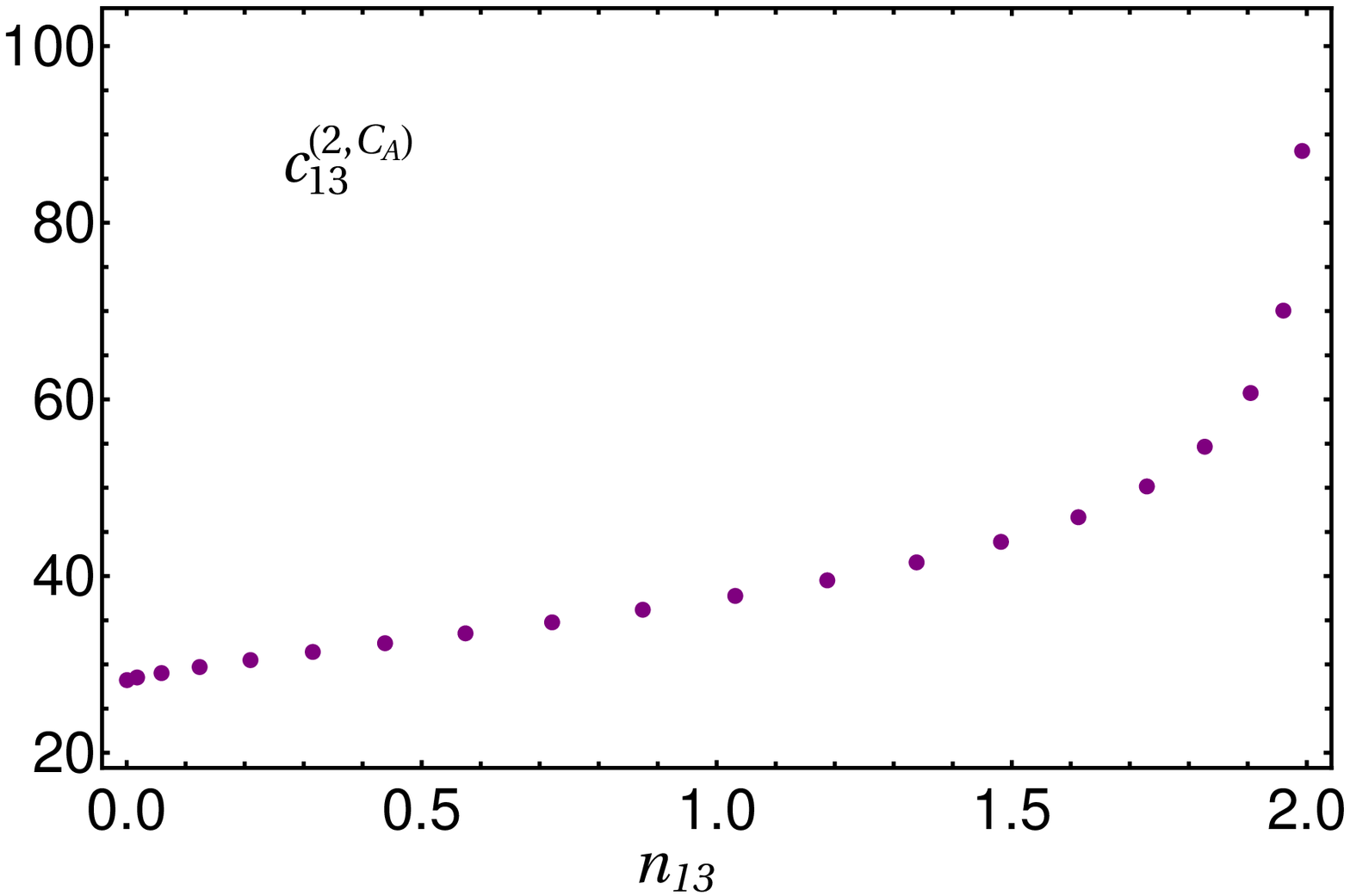}
  \\[0.5em]
 \includegraphics[width=0.31\textwidth]{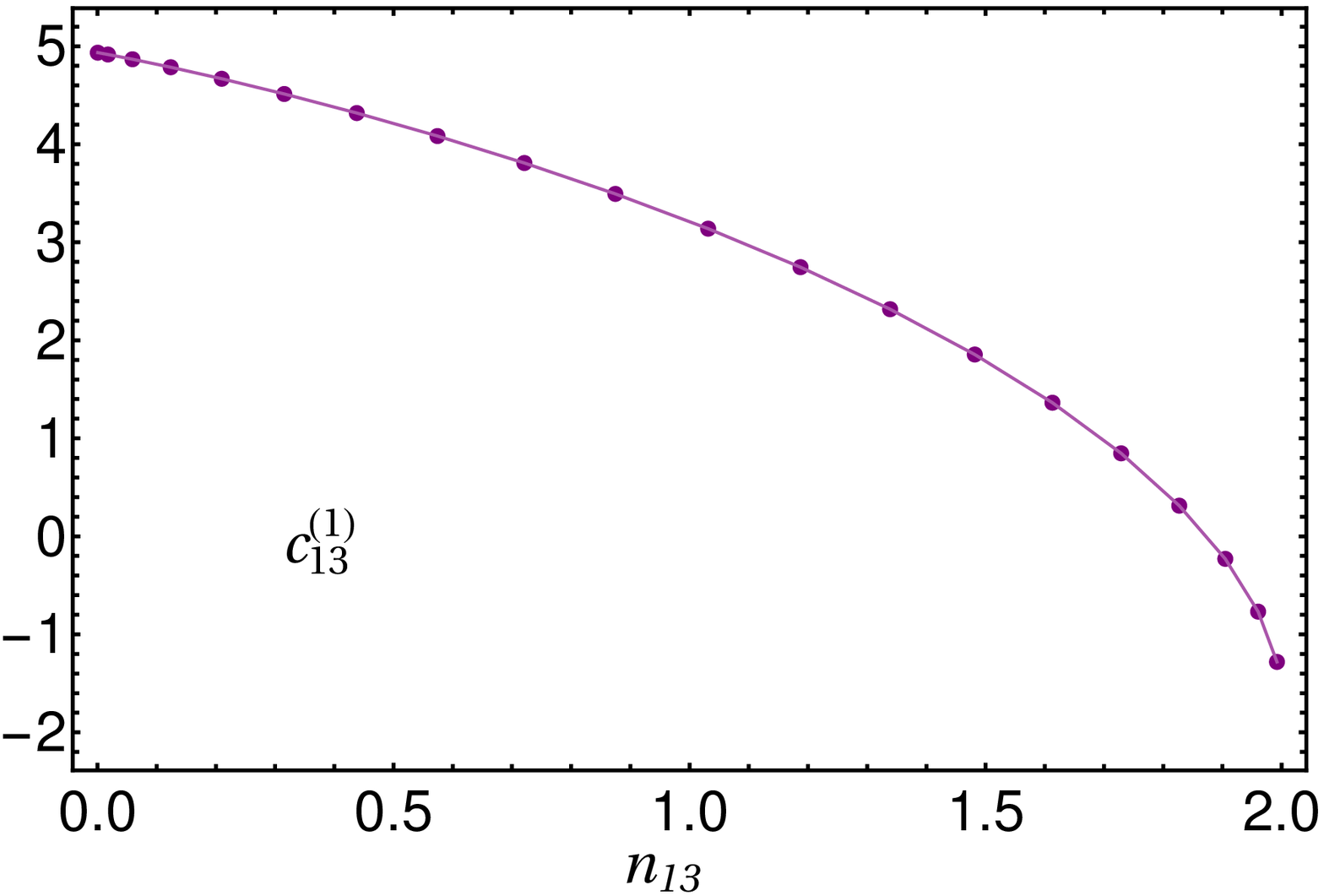}\hspace{2mm}
 \includegraphics[width=0.32\textwidth]{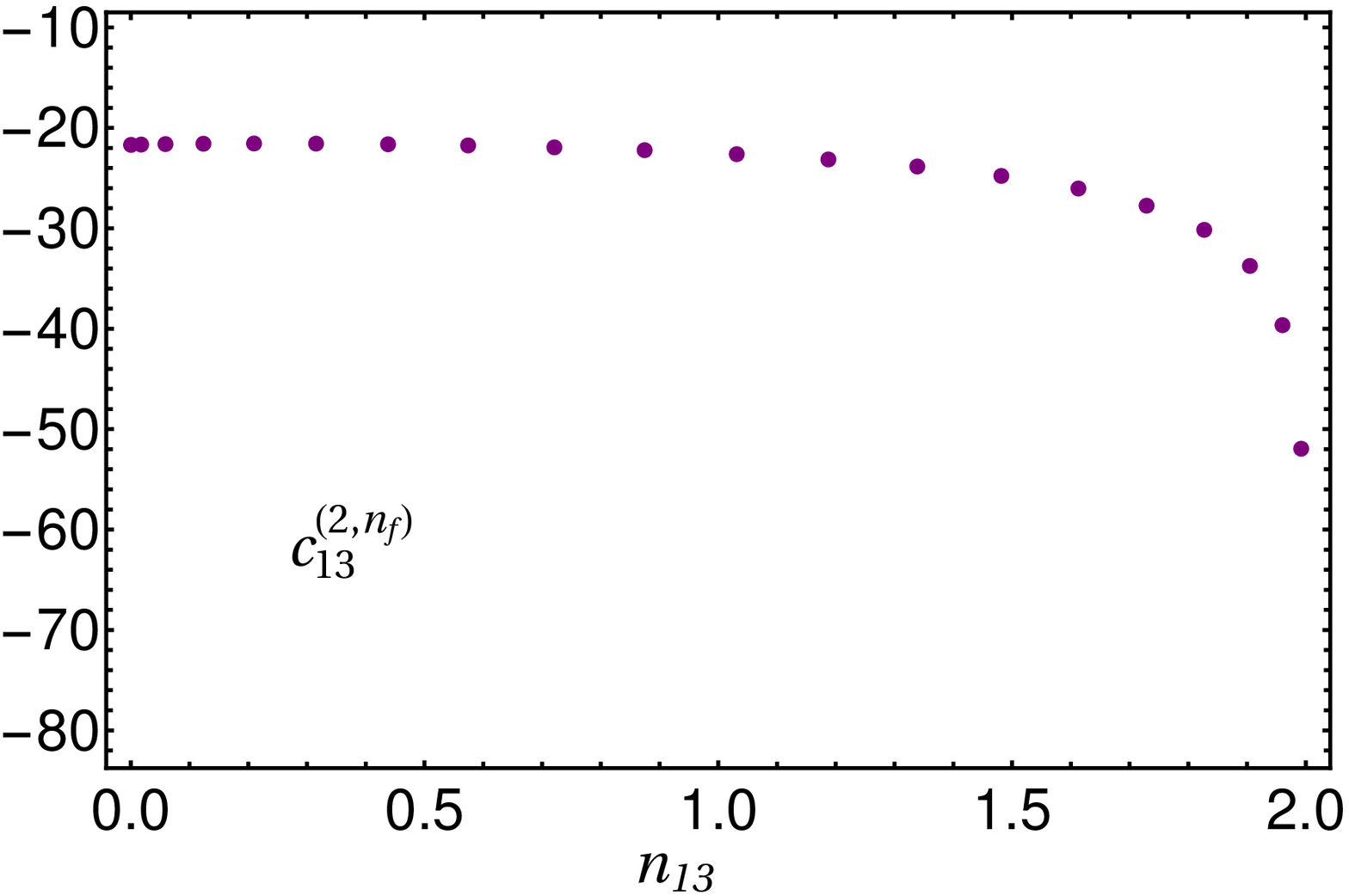}\hspace{2mm}
 \includegraphics[width=0.32\textwidth]{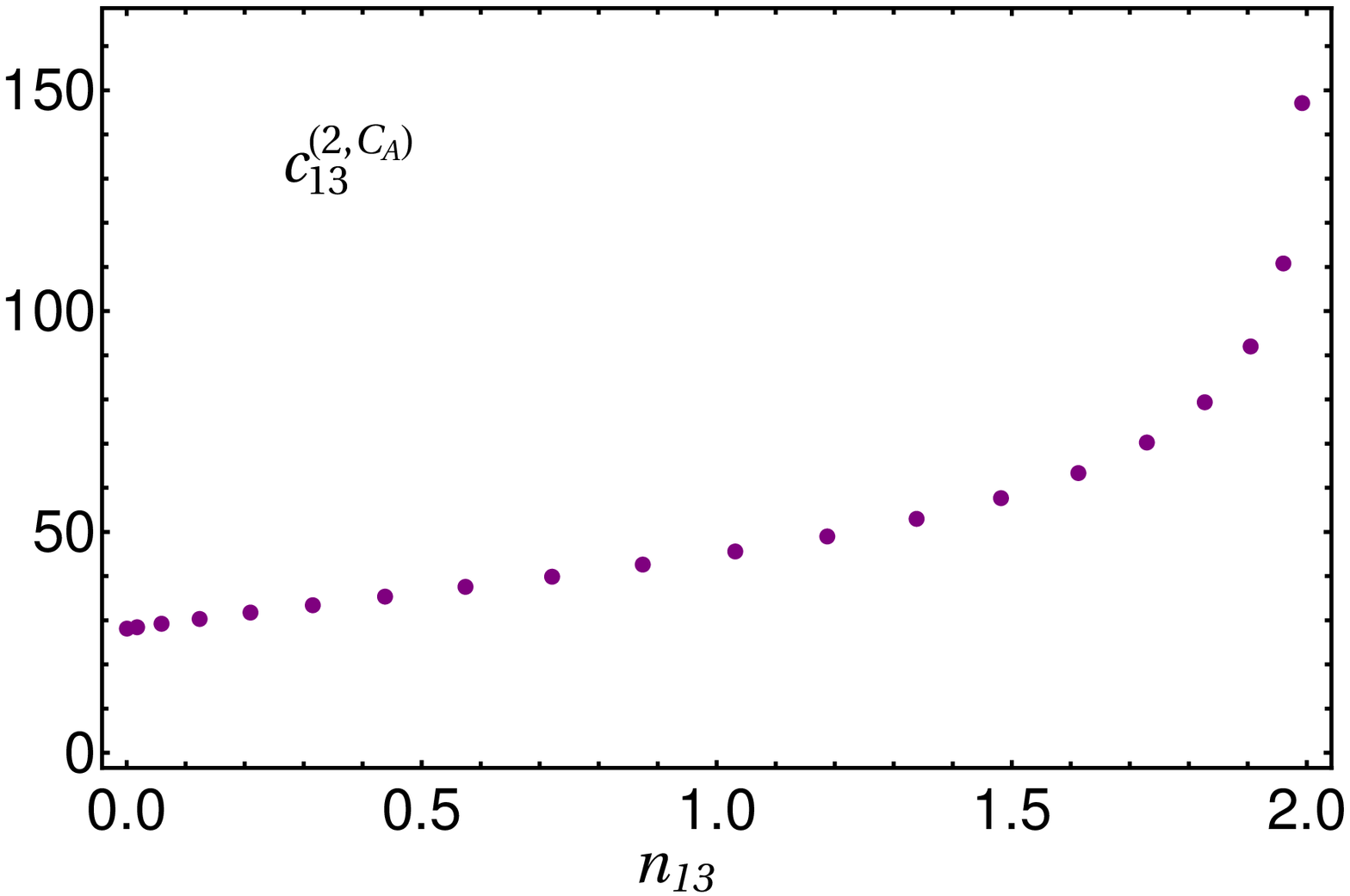}
 } 
\caption{The same as in Figure~\ref{fig:dipole:poles}, but for the finite terms of the 
renormalised soft function. The solid lines in the left plots show the result 
of the NLO calculation from~\cite{Jouttenus:2011wh}.}
\label{fig:dipole:finite}
\end{figure}

In Figure~\ref{fig:dipole:finite} we display the corresponding finite terms of the renormalised soft 
function as defined in \eqref{eq:RGE:softsolution}. The plots show the NLO coefficient 
$c_{13}^{(1)}$ (left) and the NNLO colour coefficients defined by 
$c_{13}^{(2)} = T_F \,n_f\,c_{13}^{(2,n_f)}+ C_A \,c_{13}^{(2,C_A)}$
(middle and right). Our NLO results can be compared to the calculation in~\cite{Jouttenus:2011wh}, 
which provides integral representations for the NLO dipoles for an arbitrary number of jets in distribution 
space. These results can readily be transformed to Laplace space and integrated numerically, 
which yields the solid lines in the left plots of Figure~\ref{fig:dipole:finite}. We again observe
a perfect agreement with these predictions for both the 1-jettiness and the 2-jettiness.

\begin{figure}[t!]
   \centering{ 
  \includegraphics[width=0.32\textwidth]{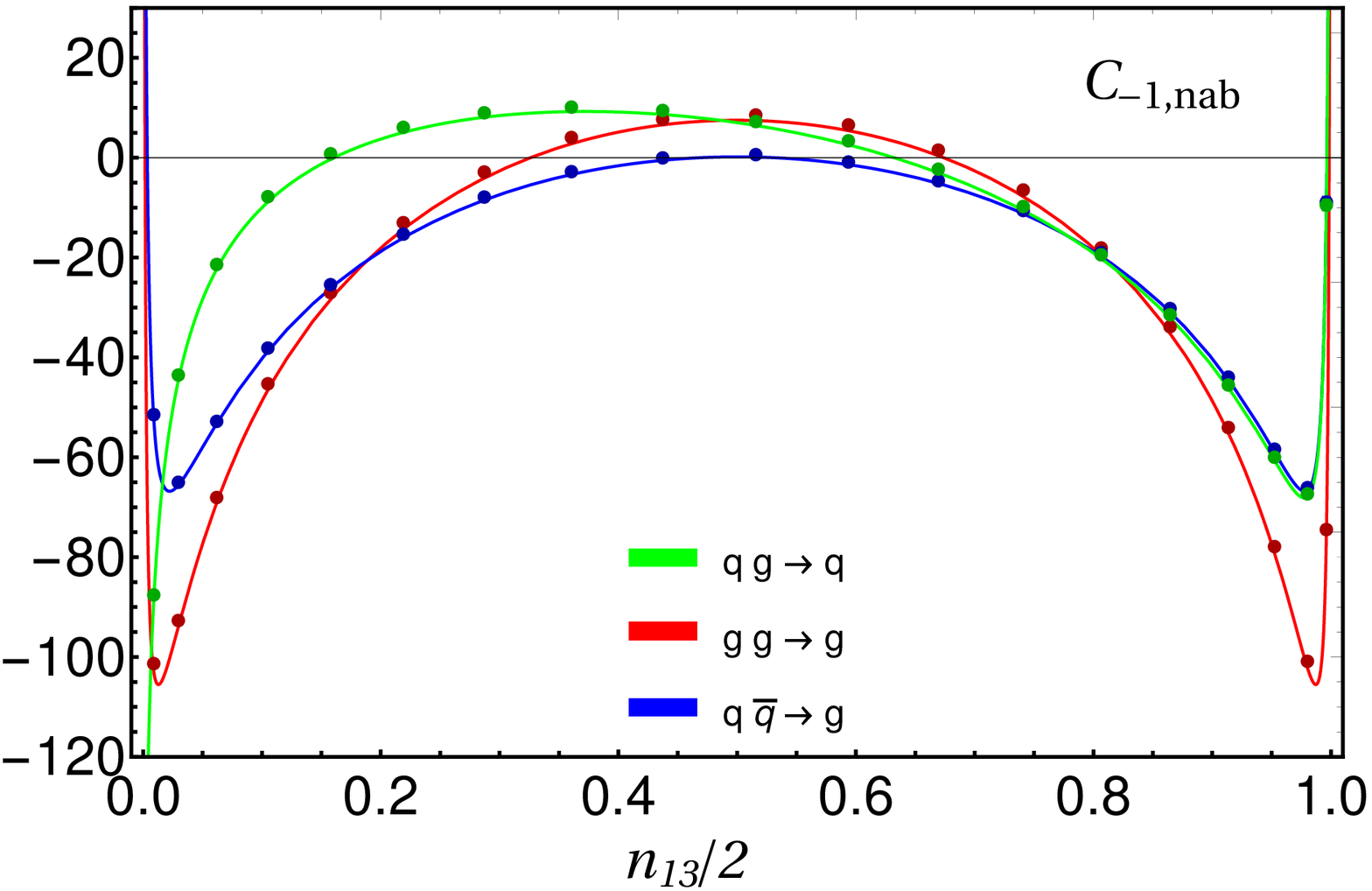}\hspace{0.cm}
  \includegraphics[width=0.32\textwidth]{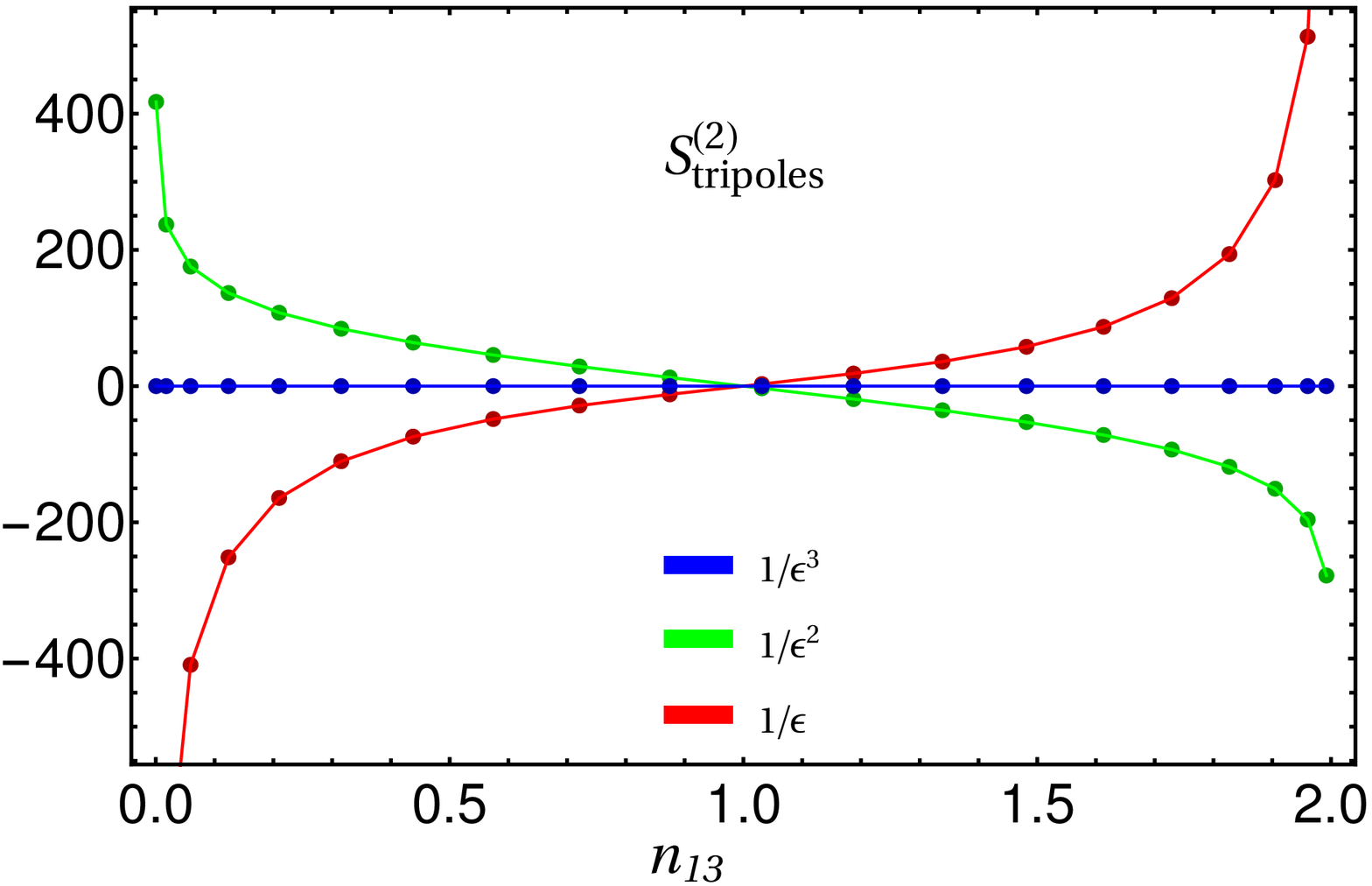}\hspace{0.cm}
  \includegraphics[width=0.32\textwidth]{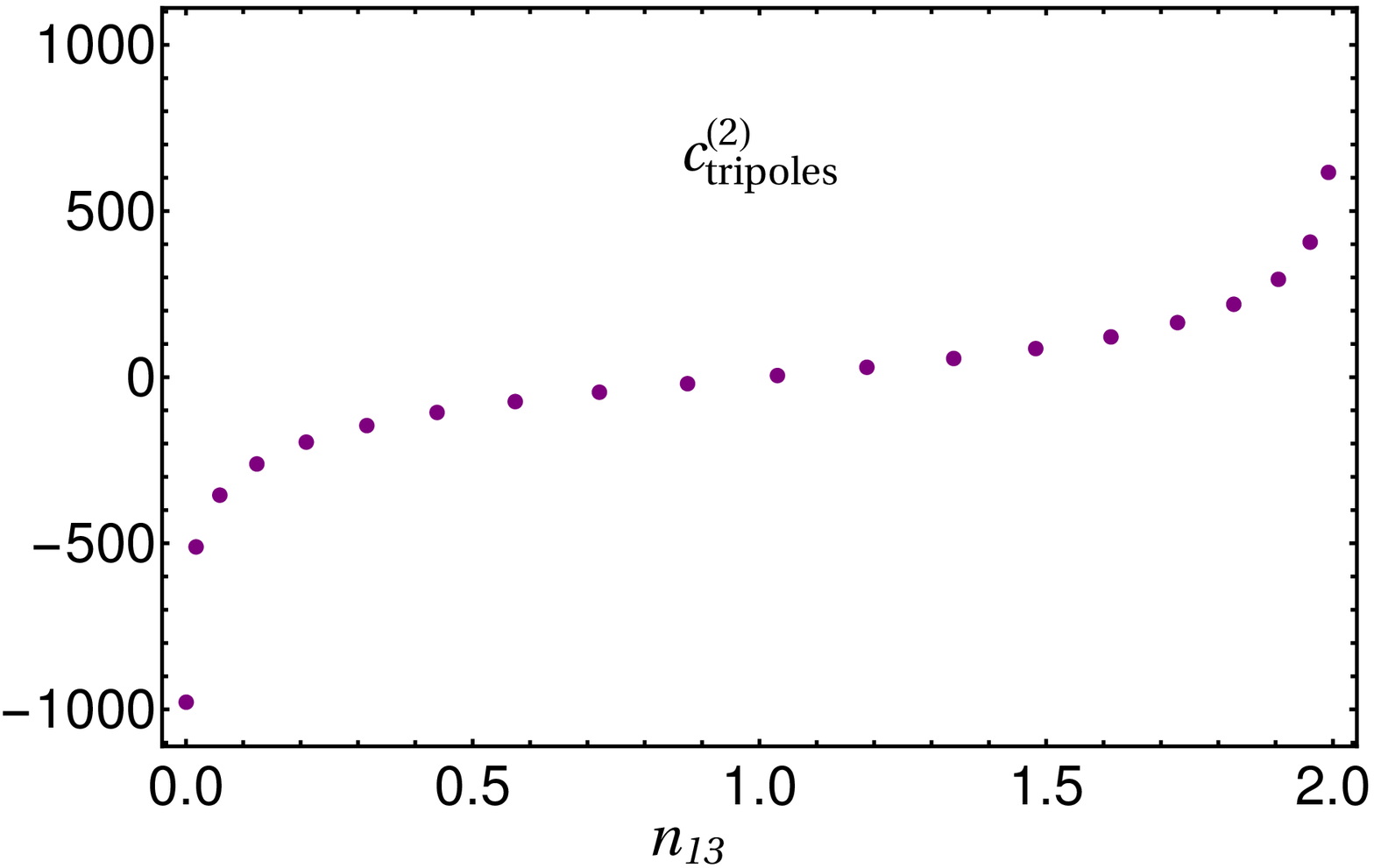}\hspace{0.cm}
}
\caption{
The first panel shows our results for the $\delta(\mathcal{T}_1)$ coefficient of the renormalised 1-jettiness
soft function (dots) in comparison with the fit functions from~\cite{Campbell:2017hsw} (solid lines). 
The second panel compares our prediction for the sum over all tripole contributions
to the 2-jettiness soft function (dots) with the one from the RG equation (solid lines).
The third panel shows the corresponding finite term of the renormalised soft function.}
\label{fig:one-jet-channel-plots}
\end{figure}

Whereas our results for the NNLO 2-jettiness soft function are new, the 1-jettiness results can be 
compared to the calculations in~\cite{Boughezal:2015eha,Campbell:2017hsw}. To this end, we adopt the
conventions used in~\cite{Campbell:2017hsw}, which provides useful fit functions for the complete NNLO 
correction for all partonic channels. We thus sum over the dipole contributions shown in 
Figure~\ref{fig:dipole:finite} with appropriate colour factors, and since the computation 
in~\cite{Campbell:2017hsw} was carried out in distribution space, we transform our results to this 
space for the comparison. The finite non-logarithmic term at NNLO is proportional to 
$\delta(\mathcal{T}_N)$ and denoted with $C_{-1,nab}$ in~\cite{Campbell:2017hsw}. Our results for this 
coefficient are shown in the first panel of Figure~\ref{fig:one-jet-channel-plots} for different partonic channels,
where the dots represent our numerical {\tt pySecDec} numbers and the solid lines are the fit functions 
from~\cite{Campbell:2017hsw}. The agreement between the two predictions is another highly non-trivial 
cross check of our calculation.

We finally address the tripole contributions, which give a non-vanishing correction only for processes
with four or more hard partons. They thus contribute to the 2-jettiness soft function, and for convenience 
we present our results for the sum over all tripole contributions in the form
\begin{equation}
\label{eq:tripoles:bare}
\sum_{i\neq j\neq k} (\lambda_{ij} - \lambda_{ip}-\lambda_{jp})\,
 f_{ABC}\; \bfT_{i}^A \;\bfT_{j}^B \;\bfT_{k}^C\;
 S_{ijk}^{(2,{\rm Im})}(\eps) \equiv
 f_{ABC}\; \bfT_{1}^A \;\bfT_{2}^B \;\bfT_{3}^C\;
 S_{\rm tripoles}^{(2)}(\eps)\,,
\end{equation}
where we have used colour conservation to bring the colour generators into this particular form. 
Our results for the divergences of the tripole contribution are represented by the dots in the 
middle panel of Figure~\ref{fig:one-jet-channel-plots}. They are again in perfect agreement with the 
predictions from the RG equation, which are indicated by the solid lines. In the right panel of 
Figure~\ref{fig:one-jet-channel-plots}, we finally display the corresponding finite term of the 
renormalised soft function in the form
\begin{align}
\sum_{i\neq j\neq k} f_{ABC}\; \bfT_{i}^A \;\bfT_{j}^B \;\bfT_{k}^C\; c_{ijk}^{(2)}  
\equiv
 f_{ABC}\; \bfT_{1}^A \;\bfT_{2}^B \;\bfT_{3}^C\;
 c_{\rm tripoles}^{(2)}\,,
\end{align}
which is another new result.

\section{Conclusions}

We presented a generalisation of a formalism that we developed earlier for the calculation of dijet
soft functions in~\cite{Bell:2015lsf,Bell:2018jvf,Bell:2018vaa}. Our method allows for a systematic
computation of NNLO soft function with an arbitrary number of light-like Wilson lines. As a first step 
to the general $N$-jet case, we focussed on SCET-1 soft functions that obey the NAE theorem.

We then applied the novel formalism to compute the $N$-jettiness soft function for single-jet and dijet 
production at hadron colliders. We checked that the poles terms of the bare soft functions agree
with the predictions from the RG equation, and we compared our NNLO 1-jettiness result with the
calculation in~\cite{Campbell:2017hsw}. Our prediction for the 2-jettiness soft function is new,
and it provides the last missing ingredient to apply the $N$-jettiness subtraction technique to
processes with two jets. As our setup is fairly general, we can also compute soft functions
for other hadronic event shapes or boosted top observables with the same techniques. Further details 
will be given in~\cite{BDMR}.

\acknowledgments

We thank Piotr Pietrulewicz for fruitful discussions and useful comparisons at the early stage 
of this work. G.B.~and B.D~are supported by the Deutsche Forschungsgemeinschaft (DFG) within 
Research Unit FOR 1873. R.R.~is supported by the Swiss National Science Foundation (SNF) under 
grant CRSII2-160814. 
Preprint numbers: SI-HEP-2018-26, QFET-2018-16.

%%%%%%%%%%%%%%%%%%%%%%%%%%%%%%%%%%%%%%%%%%%%
\bibliographystyle{JHEP}
\bibliography{SoftFunction}
%%%%%%%%%%%%%%%%%%%%%%%%%%%%%%%%%%%%%%%%%%%%

\end{document}